\documentclass[12pt,a4paper]{article}

\newcommand{\D}[2]{\frac{\partial #2}{\partial #1}}
\newcommand{\DD}[2]{\frac{\partial^2 #2}{\partial #1^2}}

\newcommand{\DDD}[2]{\frac{\partial^3 #2}{\partial #1^3}}

\renewcommand{\vec}[1]{\mbox{\boldmath$#1$}}
\newcommand{\Ord}[1]{{\cal O}\left(#1\right)}

\makeatletter
\newcounter{lineno}
\def\verbatimlisting#1{\setcounter{lineno}{0}%
    \begingroup{\footnotesize} \@verbatim \frenchspacing \@vobeyspaces
\parindent=20pt
    \everypar{\stepcounter{lineno}\llap{\thelineno\ \ }}\input#1
    \endgroup
}
\makeatother

\usepackage{url}
\usepackage{graphicx}

\newcommand{\pde}{{\textsc{pde}}}
\newcommand{\cM}{{\cal M}}
\newcommand{\ur}{u^+}
\newcommand{\ul}{u^-}
\newcommand{\up}{u_{j+1}}
\newcommand{\um}{u_{j-1}}

\newcommand{\uj}{u_{j}}
\newcommand{\xjmh}{x_{j-1/2}}
\newcommand{\xjph}{x_{j+1/2}}
\newcommand{\xjpmh}{x_{j\pm1/2}}
\newcommand{\cA}{{\cal A}}
\newcommand{\cD}{{\cal D}}

\begin{document}

\title{\sf Holistic finite differences ensure fidelity to Burger's
equation}
\author{AJ Roberts\thanks{Dept Maths \& Computing, University of
Southern Queensland, Toowoomba, Queensland 4350, Australia.
\protect\url{mailto:aroberts@usq.edu.au}}}
\maketitle

\begin{abstract}
I analyse a generalised Burger's equation to develop an accurate
finite difference approximation to its dynamics.  The analysis is
based upon centre manifold theory so we are assured that the finite
difference model accurately models the dynamics and may be constructed
systematically.  The trick to the application of centre manifold
theory is to divide the physical domain into small elements by
introducing insulating internal boundaries which are later removed.
Burger's equation is used as an example to show how the concepts work
in practise.  The resulting finite difference models are shown to be
significantly more accurate than conventional discretisations,
particularly for highly nonlinear dynamics.  This centre manifold
approach treats the dynamical equations as a whole, not just as the
sum of separate terms---it is holistic.  The techniques developed here
may be used to accurately model the nonlinear evolution of quite
general spatio-temporal dynamical systems.
\end{abstract}

\tableofcontents

\section{Introduction}

I introduce a new approach to finite difference approximation by
illustrating the concepts and analysis on the definite example of a
generalised Burger's equation.  In some non-dimensional form we
take the following partial differential equation (\pde) to govern
the evolution of $u(x,t)$:
\begin{equation}
	\D tu+\alpha u\D xu=\DD xu-\beta u^3\,.
	\label{Epde}
\end{equation}
This example equation, which includes the mechanisms of dissipation,
$u_{xx}$, and nonlinear advection/steepening, $\alpha uu_x$,
generalises Burger's equation by also including a nonlinear damping,
$\beta u^3$.  Consider implementing the method of lines by
discretising in $x$ and integrating in time as a set of ordinary
differential equations.  A finite difference approximation
to~(\ref{Epde}) on a regular grid in $x$ is straightforward, say
$x_j=jh$ for some grid spacing $h$.  For example, the linear term
\begin{displaymath}
	\DD xu = \frac{\up-2\uj+\um}{h^2}+\Ord{h^2}\,.
\end{displaymath}
However, there are differing valid alternatives for the nonlinear
term $uu_x$: two possibilities are
\begin{displaymath}
	u\D xu= \frac{\uj(\up-\um)}{2h}+\Ord{h^2}
	=\frac{\up^2-\um^2}{4h}+\Ord{h^2}\,.
\end{displaymath}
Which is better?  The answer depends upon how the discretisation of
the nonlinearity interacts with the dynamics of other terms.  The
traditional approach of considering the discretisation of each term
separately does not tell us.  Instead, in order to find the best
discretisation we have to consider the dynamics of all terms in the
equation in a holistic approach.

Centre manifold theory (see the book by Carr~\cite{Carr81}) has
appropriate characteristics to do this.  It addresses the evolution of
a dynamical system in a neighbourhood of a marginally stable fixed
point; based upon the linear dynamics the theory guarantees that an
accurate low-dimensional description of the nonlinear dynamics may be
deduced.  The theory is a powerful tool for the modelling of complex
dynamical systems
\cite[e.g.]{Carr83a,Carr83b,Meron86b,Procaccia88a,Fujimura} such as
dispersion \cite[e.g.]{Rosencrans93,Mercer94a,Watt94b}, thin fluid
films \cite[e.g.]{Chang89,Roberts94c,Roberts96b} and other
applications discussed in the review \cite{Roberts97a}.  Here we place
the discretisation of a nonlinear \pde{} such as~(\ref{Epde}) within
the purview of centre manifold theory by the following artifice (such
mathematical trickery has proven effective in thin fluid flows
\cite{Roberts94c}).  Introduce a parameter $\gamma$,
$0\leq\gamma\leq1$: at the midpoints of the grid, $x=(j+1/2)h$, insert
artificial boundaries which are ``insulating'' when $\gamma=0$, but
when $\gamma=1$ the boundaries ensure sufficient continuity to recover
the original problem.  In essence this divides the domain into
equi-sized elements centred upon each grid point, say the domain is
partitioned into $m$ elements.  For~(\ref{Epde}) we may use
\begin{equation}
	\D x\ur=\D x\ul\,,
	\quad
	(1-\gamma)\frac{h}{2}
	\left(\D x\ur+\D x\ul\right)
	=\gamma \left( \ur-\ul\right)\,,
	\label{Ebci}
\end{equation}
where $\ur$ is just to the right of a midpoint and $\ul$ to the left.
When $\gamma=1$ these reduce to conditions ensuring appropriate
continuity between adjacent elements.  When $\gamma=0$ they reduce to
conditions equivalent to the insulating
\begin{displaymath}
	\D x\ur=\D x\ul=0\,.
\end{displaymath}
We treat terms multiplied by $\gamma$ as ``nonlinear'' perturbations
to the insulated dynamics.  Then in the ``linear'' dynamics governed by
\begin{displaymath}
	\D tu=\DD xu\,,
\end{displaymath}
each element evolves exponentially quickly (in a time $\Ord{h^2}$) to
a constant value, say $u=\uj$ in the $j$th element (the particular
value depends upon the initial conditions).
But in the presence of the perturbative influences of the nonlinear
terms and the coupling between elements, the values $\uj$ associated
with each element will evolve in time.
This picture is the basis of centre manifold theory which is applied in
Section~\ref{Scm} to assure three things:
\begin{itemize}
	\item the existence of an $m$~dimensional centre manifold
	parameterized by $\uj$;

	\item the relevance of the $m$~dimensional dynamics as an accurate
	and stable model of the original dynamics~(\ref{Epde}) (sometimes
called
	asymptotic completeness~\cite{Robinson96} or~\cite{Constantin89});

	\item and that we may approximate the shape of the centre manifold and
	the evolution thereon by
	approximately solving an associated \pde.
\end{itemize}
These dynamics on the centre manifold form a finite difference
approximation.  For example, the analysis in Section~\ref{Sbe} of the
generalised Burger's equation~(\ref{Epde}) is unequivocally in favour
of the discretisation
\begin{eqnarray}
	&&
	\dot\uj +\alpha\uj\left[\frac{\up-\um}{2h}
	-\frac{\alpha \uj}{12}(\up-2\uj+\um)\right]
	\nonumber\\&&
	\approx \frac{\up-2\uj+\um}{h^2}
	\nonumber\\&&{}
	-\frac{\beta}{24}\left(\um^3 +3\um\uj^2 +16\uj^3
	+3\up\uj^2 +\up^3\right)
	\,,
	\label{Efd}
\end{eqnarray}
as an early approximation ($\dot\uj$ denotes $d\uj/dt$).  Provided the
initial conditions are not too extreme, centre manifold theory assures
us that such a discretisation models the dynamics of~(\ref{Epde}) to
errors $\Ord{\|u\|^4}$.  Observe that it is best to discretise the
nonlinear $uu_x$ terms directly, but that there is a nonlinear
correction involving the second difference.  Further, the cubic
nonlinearity is discretised in a non-obvious complicated form.
Margolin \& Jones \cite{Margolin92} have previously applied inertial
manifold ideas to discretisations of Burger's equation.  However, they
used just two basis functions on each element and invoked the
adiabatic approximation for time derivatives of ``slaved'' modes.
Here I allow arbitrarily convoluted dependence within each element and
include all effects of time variations.  The methodology presented
here provides a rigorous approach to finite difference models.

The discretisation~(\ref{Efd}) is just a low-order approximation,
centre manifold theory provides systematic corrections.  For example,
Equation~(\ref{Efd}) is obtained from terms linear in the coupling
$\gamma$.  Analysis including quadratic terms in $\gamma$ leads to the
fourth-order accurate model discussed in Section~\ref{Shoa}.  Analysis
to higher orders in the nonlinearity, discussed in
Section~\ref{Sbe}, shows higher order corrections to
the discretisation of the nonlinear terms and also incorporates
effects from a coupling of the different nonlinearities.

Computer algebra is an effective tool for modelling because of the
systematic nature of centre manifold theory
\cite{Rand85,Freire90b,Roberts96a}.  The specific finite difference
models presented here were derived by the computer algebra program
given in Appendix~\ref{Sca}.  Such a computer algebra program may
be straightforwardly modified to model more complicated dynamical
systems.

In this work I concentrate upon a proof of the concept of applying
centre manifold theory to constructing effective finite difference
models.  To that end we only consider an infinite domain or strictly
periodic solutions in finite domains.  Then all elements of the
discretisation are identical by symmetry and the analysis of all
elements is simultaneous.  However, if physical boundaries to the
domain of the \pde{} are present, then those elements near the
physical boundary will need special treatment.  I do not see the need
for any new concepts, just an increase in the amount of detail.  The
centre manifold approach also sheds an interesting light upon the
issue of the initial condition for a finite difference approximation.
Earlier work on the issue of initial conditions in general
\cite{Roberts89b,Cox93b,Roberts97b} hints that the initial values for
the parameters $u_j(t)$ is not simply the field $u$ sampled at the
grid points, $u(x_j,0)$, but some more subtle transform.  Some
preliminary research suggests that a leading order approximation is
that $u_j(0)$ is the element average of $u(x,0)$, an approximation
which usefully conserves $u$.  Lastly, here we have only analysed an
autonomous dynamical system, the \pde~(\ref{Epde}); forcing applied to
the \pde{} may be approximated using the projection obtained for
initial conditions \cite{Cox91,Roberts97b}.  Further research is
needed in the above issues and in the application of the theory to
higher order \pde{}s and in higher spatial dimensions.

\section{Centre manifold theory underpins the fidelity}
\label{Scm}

Here I describe in detail one way to place the discretisation of
\pde{}s within the purview of centre manifold theory.  For
definiteness I address the generalised Burger's equation~(\ref{Epde})
as an example of a broad class of \pde{}s.

As introduced earlier, the discretisation is established via an
equi-spaced grid of collocation points, $x_j=jh$ say, for some
constant spacing $h$.  At midpoints $x_{j+1/2}=(x_j+x_{j+1})/2$
artificial boundaries are introduced with one extra refinement over
that discussed in the Introduction:
\begin{eqnarray}
	&&
	\D x\ur=\D x\ul\,,\label{Ebcsd}
	\\&&
	(1-\gamma)\cA\frac{h}{2}\left(\D x\ur
	+\D x\ul\right)
	=\gamma \left( \ur-\ul\right)\,,
	\label{Esbci}
\end{eqnarray}
where the introduction of the near identity operator
\begin{equation}
	\cA
	=1-\frac{h^2}{12}\D t{}+\frac{h^4}{120}\DD t{}
	-\frac{17h^6}{20160}\DDD t{}+\cdots\,,
	\label{Efudge}
\end{equation}
will be explained in the next section.  These boundaries divide the
domain into a set of elements, the $j$th element centred upon $x_j$
and of width $h$.  A non-zero value of the parameter $\gamma$ couples
these elements together so that when $\gamma=1$ the \pde{} is
effectively restored over the whole domain.  The generalised Burger's
equation~(\ref{Epde}) with ``internal boundary
conditions''~(\ref{Ebcsd}--\ref{Esbci}) is analysed here to give the
discretisation in the interior of the domain.

The application of centre manifold theory is based upon a linear
picture of the dynamics.  Adjoin the dynamically trivial equation
\begin{equation}
	\D t\gamma=0\,,
	\label{Etriv}
\end{equation}
and consider the dynamics in the extended state space $(u(x),\gamma)$.
This is a standard trick used to unfold bifurcations
\cite[\S1.5]{Carr81} or to justify long-wave approximations
\cite{Roberts88a}.  Within each element $u=\gamma=0$ is a fixed point.
Linearized about each fixed point, that is to an error
$\Ord{\|u\|^2+\gamma^2}$, the \pde{} is
\begin{displaymath}
	\D tu=\DD xu\,,
	\quad\mbox{s.t.}\quad
	\cA\left.\D xu\right|_{x=\xjpmh}=0\,,
\end{displaymath}
namely the diffusion equation with essentially insulating boundary
conditions.  There are thus linear eigenmodes associated with each
element:
\begin{equation}
\gamma=0\,,\quad
	u\propto \left\{
	\begin{array}{ll}
		e^{\lambda_nt}\cos[n\pi(x-\xjmh)/h]\,, & \xjmh<x<\xjph\,,  \\
		0\,, & \mbox{otherwise}\,,
	\end{array}\right.
	\label{Emode}
\end{equation}
for $n=0,1,\ldots$, where the decay rate of each mode is
\begin{equation}
	\lambda_n=-\frac{n^2\pi^2}{h^2}\,;
	\label{Eeigen}
\end{equation}
together with the trivial mode $\gamma=\mbox{const}$, $u=0$.  In a
domain with $m$ elements, evidentally all eigenvalues are negative,
$-\pi^2/h^2$ or less, except for $m+1$ zero eigenvalues: $1$
associated with each of the $m$ elements and $1$ from the
trivial~(\ref{Etriv}).  Thus, provided the nonlinear terms
in~(\ref{Epde}) are sufficiently well behaved, the existence theorem
(\cite[p281]{Carr83b} or~\cite[p96]{Vanderbauwhede89}) guarantees that
a $m+1$~dimensional centre manifold $\cM$ exists for~(\ref{Epde})
with~(\ref{Ebcsd}--\ref{Etriv}).  The centre manifold $\cM$ is
parameterized by $\gamma$ and a measure of $u$ in each element, say
$\uj$: using $\vec u$ to denote the collection of such parameters,
$\cM$ is written as
\begin{equation}
	u(x,t)=v(x;\vec u,\gamma)\,.
	\label{Ecmv}
\end{equation}
In this the analysis has a very similar appearance to that of finite
elements.  The theorem also asserts that on the centre manifold the
parameters $\uj$ evolve deterministically
\begin{equation}
	\dot\uj=g_j(\vec u,\gamma)\,,
	\label{Ecmg}
\end{equation}
where $g_j$ is the restriction of~(\ref{Epde})
and~(\ref{Ebcsd}--\ref{Etriv}) to $\cM$.  In this approach the
parameters of the description of the centre manifold may be anything
that sensibly measures the size of $u$ in each element---I simply
choose the value of $u$ at the grid points, $\uj=u(x_j,t)$.  This
provides the necessary amplitude conditions, namely that
\begin{equation}
	\uj=v|_{x_j}\,.
	\label{Eamp}
\end{equation}
The above application of the theorem establishes that in principle we
may find the dynamics~(\ref{Ecmg}) of the interacting elements of the
discretisation.  A low order approximation is given in~(\ref{Efd}).

The next outstanding question to answer is: how can we be sure that
such a description of the interacting elements does actually
\emph{model} the dynamics of the original system~(\ref{Epde})
with~(\ref{Ebcsd}--\ref{Etriv})?  In the development of inertial
manifolds by Temam \cite{Temam90} and others, this question is
sometimes phrased as one about the asymptotic completeness of the
model, for example see Robinson \cite{Robinson96} or Constantin et al
\cite[Chapt.12--3]{Constantin89}.  Here, the relevance theorem of
centre manifolds, \cite[p282]{Carr83b}
or~\cite[p128]{Vanderbauwhede89}, guarantees that all solutions
of~(\ref{Epde}) and~(\ref{Ebcsd}--\ref{Etriv}) which remain in the
neighbourhood of the origin in $(u(x),\gamma)$ space are exponentially
quickly attracted to a solution of the $m$ finite difference
equations~(\ref{Ecmg}).  For practical purposes the rate of attraction
is estimated by the leading negative eigenvalue, here
$-\pi^2/h^2$.  Centre manifold theory also guarantees that the
stability near the origin is the same in both the model and the
original.  Thus the finite difference model will be stable if the
original dynamics are stable.  After exponentially quick transients
have died out, the finite difference equation~(\ref{Ecmg}) on the
centre manifold accurately models the complete system~(\ref{Epde})
and~(\ref{Ebcsd}--\ref{Etriv}).

The last piece of theoretical support tells us how to approximate the
shape of the centre manifold and the evolution thereon.  Approximation
theorems such as that by Carr \& Muncaster \cite[p283]{Carr83b} assure
us that upon substituting the ansatz~(\ref{Ecmv}--\ref{Ecmg}) into the
original~(\ref{Epde}) and~(\ref{Ebcsd}--\ref{Etriv}) and solving to
some order of error in $\|u\|$ and $\gamma$, then $\cM$ and the
evolution thereon will be approximated to the same order.  The catch
with this application is that we need to evaluate the approximations
at $\gamma=1$ because it is only then that the artificial internal
boundaries are removed.  In some applications of such an artificial
homotopy I have demonstrated good convergence in the parameter
$\gamma$ \cite{Roberts94c}.  Thus although the order of error
estimates do provide assurance, the actual error due to the evaluation
at $\gamma=1$ should be also assessed otherwise.  Here, as discussed
in Section~\ref{Shoa}, we have crafted the interaction~(\ref{Esbci})
between elements so that low order terms in $\gamma$ recover the exact
finite difference formula for linear terms.  Note that although centre
manifold theory ``guarantees'' useful properties near the origin in
$(u(x),\gamma)$ space, because of the need to evaluate asymptotic
expressions at $\gamma=1$, I have used a weaker term elsewhere, namely
``assures''.

\section{Numerical comparisons show the effectiveness}
\label{Sbe}

We now turn to a detailed description of the centre manifold model for
Burger's equation~(\ref{Epde}).

The algebraic details of the derivation of the centre manifold
model~(\ref{Ecmv}--\ref{Ecmg}) is the task of the computer algebra
program listed in the Appendix.  In an algorithm introduced
in~\cite{Roberts96a}, the program iterates to drive to zero the
residuals of the governing differential equation~(\ref{Epde}) and its
boundary conditions~(\ref{Ebcsd}--\ref{Esbci}).  A key part of the
iteration is to solve for corrections $v'$ and $g'$ from the linear
diffusion equation within each element
\begin{displaymath}
	\DD x{v'}=g'+R\,,\quad
	\left.\D x{v'}\right|_{\xjpmh}=R_\pm\,,
\end{displaymath}
where $R$ and $R_\pm$ denote the residuals for the current
approximation.  The initial approximation is simply that in each
element $u=u_j$, constant, with $\dot u_j=g_j=0$.  The algebraic
details of this iteration are largely immaterial so long as the
residuals iterate to zero to some order in $u$ and $\gamma$.  This is
achieved by the listed computer algebra program, it is included to
replace the recording of tedious details of elementary algebra.

\begin{figure}[tbp]
	\centering
\includegraphics[width=0.9\textwidth]{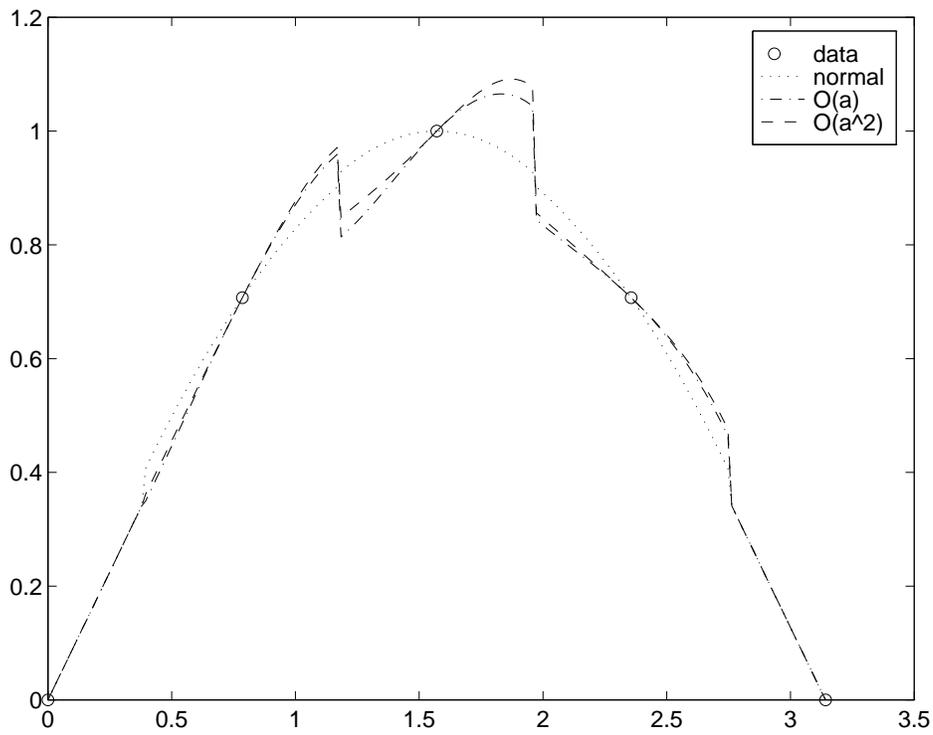} \caption{\narrower low
order
approximations~(\ref{Ecm23}) to the centre manifold field $u(x)$ within
each element for a fixed set of data points (circles) corresponding to
$m=8$ intervals on $[0,2\pi)$ with parameters $\alpha=6$, $\beta=0$ and
$\gamma=1$: $\cdots$, the normal field associated with finite
differences; $-\cdot-\cdot-$, includes
the first effects of the nonlinear advection term, $\alpha uu_x$;
$-~-~-$, includes the second-order effects in $\alpha$.  }
	\label{Fsine}
\end{figure}
The shape of the centre manifold gives the field $u$ as a function of
the parameters $\uj$ and the coupling $\gamma$.  To low-order in $u$
and $\gamma$, and written in terms of the scaled coordinate
$\xi=(x-x_j)/h$, the solution field in the $j$th element is
\begin{eqnarray}
	u & = & \uj
	+\gamma\left[ \frac{\up-\um}{2}\xi
	+\frac{\up-2\uj+\um}{2}\xi^2 \right]
	\nonumber\\&&{}
	+h\gamma\left[\alpha \uj(\up-2\uj+\um)\left(
	\frac{1}{6}\xi^3-\frac{1}{8}\xi \right)\right]
	\nonumber\\&&{}
	+h^2\gamma\left[ \alpha^2 \uj^2(\up-2\uj+\um)\left(
	\frac{1}{24}\xi^4-\frac{1}{48}\xi^2 \right)
	\right.\nonumber\\&&\quad\left.{}
	+\beta (\up^3-3\up\uj^2+4\uj^3-3\um\uj^2+\um^3)
	\left(-\frac{1}{24}\xi^4-\frac{1}{48}\xi^2\right)
	\right.\nonumber\\&&\quad\left.{}
	+\beta (\up^3-3\up\uj^2+3\um\uj^2-\um^3)
	\left(-\frac{1}{12}\xi^3+\frac{1}{48}\xi\right)
	\right.\nonumber\\&&\quad\left.{}
	-\beta\frac{\uj^2}{8}\left( (\up-\um)\xi+(\up-2\uj+\um)\xi^2 \right)
	\right]
	\nonumber\\&&{}
	+\Ord{\|u\|^4,\gamma^2}\,.
	\label{Ecm23}
\end{eqnarray}
Observe the first line, when evaluated at $\gamma=1$, is simply the
quadratic interpolant based upon second order accurate centred
differences.  This is normal finite differences.  As displayed in
Figure~\ref{Fsine}, the second line shows that this field should be
modified because of the nonlinear advection term $\alpha uu_x$; the
modification in proportion to the second difference at $x_j$ is
reasonable because when $u$ has a local maximum the field must be
decreased/increased to the left/right due to the advection to the
right of the lower/higher levels of $u$ respectively.  Note that the
same discretisation for this nonlinear term is obtained here whether
we code it as $\alpha uu_x$ or $\alpha(u^2)_x/2$---the centre manifold
is independent of any valid change in the algebraic description of the
dynamics.  The third line gives the next order correction due to the
nonlinear advection and is also displayed in Figure~\ref{Fsine}.  The
fourth, fifth and sixth lines show how the cubic nonlinearity, $\beta
u^3$, modifies $u$ in the element because of its nonlinearly varying
effect as a source or sink when the field $u$ itself varies.  It might
appear odd to introduce the dysfunctional behaviour between elements
shown in Figure~\ref{Fsine}, but centre manifold theory reasonably
assures us that it is appropriate for the nonlinear dynamics we wish
to model.  In short, the description of the centre manifold is based
upon standard differencing formulae, but includes effects upon $u$
within each element due to the nonlinear processes that act in the
continuum dynamics.

The finite difference model is given by the evolution on the centre
manifold.  To linear terms in $\gamma$ but to two orders higher
in $\|u\|$ it is
\begin{eqnarray}
	\dot\uj & = &
	\frac{\gamma}{h^2}\left( \up-2\uj+\um \right)
	-\frac{\gamma\alpha}{2h}\uj(\up-\um)
	-\beta\uj^3
	\nonumber\\&&{}
	+\gamma\left[ \frac{\alpha^2}{12}\uj^2(\up-2\uj+\um)
	\right.\nonumber\\&&\left.\quad{}
	+\frac{\beta}{24}\left(-\up^3-3\up\uj^2+8\uj^3-3\um\uj^2-\um^3\right) \right]
\nonumber\\&&{}
	+h\gamma\left[ \frac{\alpha\beta}{8}\uj^3
	(\up-\um) \right]
	\nonumber\\&&{}
	+h^2\gamma\left[
	-\frac{\alpha^4}{720}\uj^4(\up-2\uj+\um)
	\right.\nonumber\\&&\left.\quad{}
	+\frac{\alpha^2\beta}{5760}\uj^2\left( 9\up^3 -113\up\uj^2 +208\uj^3
	-113\um\uj^2 +9\um^3 \right)
	\right.\nonumber\\&&\left.\quad{}
	+\frac{\beta^2}{1920}\left( -21\up^5 +6\up^3\uj^2 +7\up\uj^4 +16\uj^5
	\right.\right.\nonumber\\&&\left.\left.\qquad{}
	+7\um\uj^4 +6\um^3\uj^2 -21\um^5 \right)
	\phantom{\frac11}\right]
	\nonumber\\&&{}
	+\Ord{\|u\|^6,\gamma^2}\,.
	\label{Ecmg24}
\end{eqnarray}
The first line recorded here, when evaluated for $\gamma=1$, is
just the classic second-order finite difference equation for the
generalised Burger's equation~(\ref{Epde}).  The second line starts
accounting systematically for the variations in the field $u$
within each element and how they affect the evolution through the
nonlinear terms.  The approximation formed by the first three lines
is that reported in the Introduction as~(\ref{Efd}).  The fourth
and further lines above, through $\alpha\beta$ and $\alpha^2\beta$
effects, show that this approach also accounts for interaction
between the nonlinear terms in the \pde{} in the finite difference
approximation.  Finite difference equations derived via this
approach holistically model all the interacting dynamics of the
entire \pde.

\begin{figure}[tbp]
	\centering
	\includegraphics[width=0.9\textwidth]{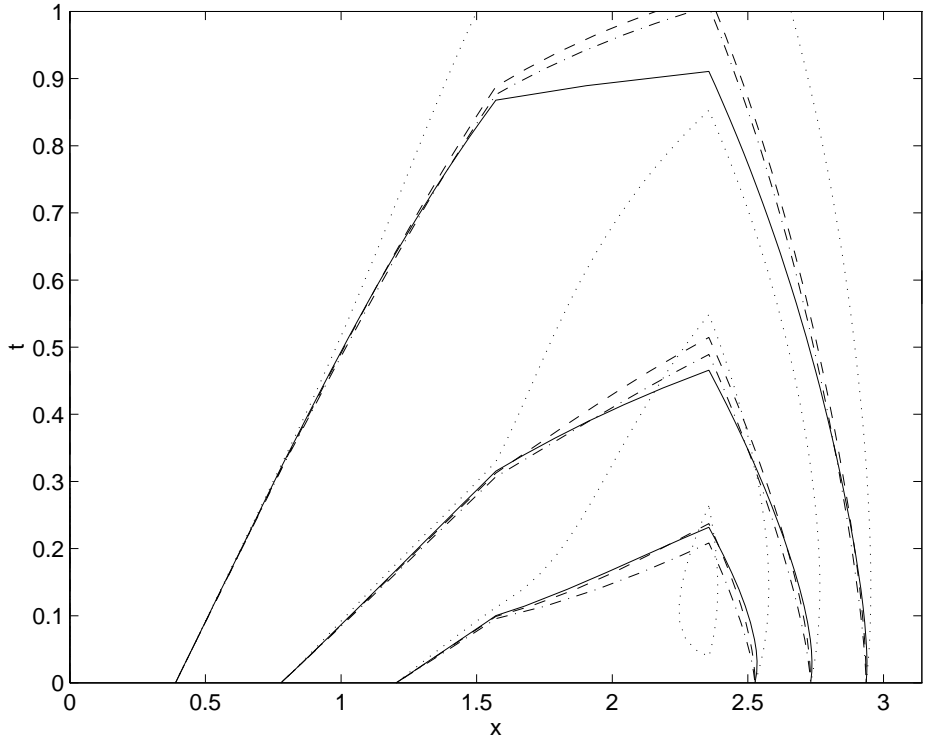} \caption{Contours
	of an exact solution $u(x,t)$, -----, to compare with numerical
	approximations~(\ref{Eburg2}): $\cdots$, the conventional
	approximation, errors $o(\alpha)$; $-\cdot-\cdot-$, the first
	correction, errors $o(\alpha^2)$; $-~-~-$, the second correction,
	errors $o(\alpha^4)$.  Burger's equation~(\ref{Epde}) with
	parameters $\alpha=6$ and $\beta=0$ is discretised on $m=8$
	elements in $[0,2\pi)$ and drawn with contour interval $\Delta
	u=0.2$.}
	\label{F286}
\end{figure}
To show the effectiveness of the approach I compare the finite
difference model~(\ref{Ecmg24}) to exact solutions of Burger's
equation~(\ref{Epde}) with $\beta=0$.  Exact solutions are obtained
via the Cole-Hopf transformation $u=-(2/\alpha)\phi_x/\phi$ by
choosing $\phi(x,t)$ satisfying the diffusion equation
$\phi_t=\phi_{xx}$.  For example, here I choose $2\pi$-periodic
functions
\begin{displaymath}
	\phi=\frac{14}{\sqrt{t_0}}e^{-\pi^2/4t_0}
	+\sum_k\frac{1}{\sqrt{t+t_0}}
	\exp\left[ -\frac{(x-k2\pi)^2}{4(t+t_0)} \right]\,,
\end{displaymath}
with $t_0=\pi/(4\sqrt \alpha)$, chosen so that the initial state
$u(x,0)$ is a rough approximation to $\sin(x)$. Note that
consequently $\|u(x,0)\|\approx 1$.  Choosing $m$
intervals on $[0,2\pi)$ gives an element length $h=2\pi/m$ and grid
points $x_j=jh$ for $j=0,\ldots,m-1$.  Because of the antisymmetry
in $u(x,t)$ about $x=k\pi$, I only display the interval $[0,\pi]$.
With $\beta=0$, there are three different approximations
from~(\ref{Ecmg24}), with $\gamma=1$, depending upon where the
expansion is truncated in $\|u\|$ (or equivalently in $\alpha$):
\begin{eqnarray}
	\dot\uj & \approx &
	\frac{1}{h^2}\left( \up-2\uj+\um \right)
	-\frac{\alpha}{2h}\uj(\up-\um)
	\nonumber\\&&{}
	+\frac{\alpha^2}{12}\uj^2(\up-2\uj+\um)
	\nonumber\\&&{}
	-h^2\frac{\alpha^4}{720}\uj^4(\up-2\uj+\um)\,.
	\label{Eburg2}
\end{eqnarray}
Just the first line forms a model with $o(\alpha)$ errors (a normal
finite difference approximation), the first two lines form a model
with $o(\alpha^2)$ errors, and all shown terms form the model with
$o(\alpha^4)$ errors.  The solutions of these models over $0<t<1$
with $m=8$ and nonlinearity parameter $\alpha=6$ are shown in
Figure~\ref{F286}.  Observe that the leading approximation (dotted)
is significantly in error whereas the next two refinements
(dot-dashed and dashed) are reasonably accurate.  Such accuracy is
remarkable considering the high level of nonlinearity, $\alpha=6$,
and the few points in the discretisation, $m=8$.

\begin{table}[tbp]
	\centering
\caption{Comparison of the errors $\varepsilon$, defined
in~(\ref{Eerr}), of the spatially second order model~(\ref{Eburg2}) at
the two first truncations in the nonlinear parameter $\alpha$ for
various grid subintervals $m$.  Observe that the higher order
model is significantly less sensitive to the nonlinearity.}
\begin{tabular}{|r|rr|rr|rr|}
	\hline
	 & \multicolumn{2}{c|}{$m=8$}  & \multicolumn{2}{c|}{$m=16$}  &
	 \multicolumn{2}{c|}{$m=32$}   \\
	$\alpha$ & $o(\alpha)$ & $o(\alpha^2)$ & $o(\alpha)$ & $o(\alpha^2)$ &
$o(\alpha)$ & $o(\alpha^2)$ \\
	\hline\hline
	1 & .0116 & .0095 & .0031 & .0025 & .0008 & .0006  \\
	\hline
	3 & .0356 & .0109 & .0102 & .0040 & .0026 & .0011  \\
	\hline
	6 & .0694 & .0115 & .0215 & .0059 & .0054 & .0018  \\
	\hline
	10 & .0971 & .0186 & .0318 & .0055 & .0081 & .0026  \\
	\hline
\end{tabular}
	\label{Ta2}
\end{table}
For an assessment of errors over a range of parameters I record in
Table~\ref{Ta2} the error
\begin{equation}
	\varepsilon=\max_t\left(\frac{1}{m}\sum_j|\epsilon_j(t)|\right)\,,
	\label{Eerr}
\end{equation}
where $\epsilon_j(t)$ is the difference between an approximation
and the exact solution at the $j$th grid point.  Observe the usual
properties that the errors decrease with increasing spatial
resolution $m$, and that the errors increase with increasing
nonlinearity $\alpha$.  Our interest lies more in the comparison
between the $o(\alpha)$ and $o(\alpha^2)$ models (here the
$o(\alpha^4)$ model is virtually indistinguishable from the
$o(\alpha^2)$ model).  For small nonlinearity, $\alpha=1$, there is
very little difference; presumably the errors are dominated by the
second-order approximation of spatial derivatives.  As the
nonlinearity $\alpha$ increases, the error of the $o(\alpha)$
approximation increases roughly in proportion to $\alpha$, but the
error of the $o(\alpha^2)$ approximation grows much more slowly.
We conclude, as centre manifold theory assures us, that the
nonlinearly corrected $o(\alpha^2)$ model more accurately captures
the nonlinear dynamics of Burger's equation.

\section{Higher order approximations are more accurate}
\label{Shoa}

So far I have reported results only to first order in the coupling
coefficient $\gamma$.  Retaining higher orders in $\gamma$ gives
higher order difference rules as the coupling between adjacent
elements is always ameliorated by a power of $\gamma$.  However, it
is only with the specific coupling of~(\ref{Esbci}) that the $2p+1$
width stencil, obtained by retaining terms in $\gamma^p$, attains
$2p$th-order accuracy in the linear terms.  Retaining $\gamma^2$
terms, for example, then gives a fourth-order model which also
performs remarkably well, especially for larger nonlinearities.

The particular form of the artificial internal boundary
condition~(\ref{Esbci}) controls the flow of information between
elements of the discretisation.  When $\gamma=1$ the condition reduces
to requiring the desired continuity: $\ul=\ur$ from the rightmost term
and $\ul_x=\ur_x$ from~(\ref{Ebcsd}) where, as before, the superscripts
$\pm$ denote evaluation at the internal boundary of expressions from
the right/left-hand elements respectively.  But we have enormous
freedom in choosing the $(1-\gamma)$-term in~(\ref{Esbci}).  Our first
requirement is that when $\gamma=0$ it insulates the elements from
each other so that the values of $\uj$ in each element are independent
in the centre manifold analysis.  This is achieved is conjunction
with~(\ref{Ebcsd}) by only involving $\ul_x+\ur_x$ and its time
derivatives.  This is all that is necessary to apply centre
manifold theory.

However, computer algebra experiments show that generally we have to
sum to high-order in $\gamma$ to recover standard finite difference
formulae for the linear diffusion term $u_{xx}$.  This is impractical
because the width of the finite difference formula also grows with the
order of $\gamma$.  But we have freedom to find an interaction
where the expansions of the linear terms truncate in $\gamma$ and
thus recover the standard formulae at low-orders in $\gamma$.
Computer algebra experiments show that the expansion~(\ref{Efudge})
for $\cA$ ensures the linear diffusion term is modelled with errors
$\Ord{h^{2p}}$ by a $2p+1$ width finite difference stencil when the
centre manifold is constructed with errors $\Ord{\gamma^{p+1}}$, for
$p=1,\ldots,4$.  The expression for $\cA$ in~(\ref{Efudge}) is
recognised to be the ``asymptotic expansion'' of the operator
\begin{displaymath}
	\cA
	=\frac{2}{h\sqrt{\partial_t}}\tanh\left( h\sqrt{\partial_t}/2
	\right)\,.
\end{displaymath}
It is not apparent why this particular operator is so desirable.
However, an answer may lie via the following observation.  In the
linear dynamics, $u_t=u_{xx}$ and so effectively
$\partial_t=\partial_x^2$.  Denoting the fundamental spatial
differential operator $\cD=h\partial_x$, observe that in effect
\begin{displaymath}
	\cA=\frac{2}{\cD}\tanh(\cD/2)=\frac{\delta}{\cD\mu}\,,
\end{displaymath}
where $\delta$ denotes the centred difference operator and $\mu$ the
centred averaging operator:
\begin{displaymath}
	\delta\uj=u_{j+1/2}-u_{j-1/2}\,,\quad
	\mu\uj=\frac{1}{2}\left( u_{j+1/2}+u_{j-1/2} \right)\,.
\end{displaymath}
Using this in~(\ref{Efudge}), ``cancelling'' $\cD$ and
``multiplying'' by $\mu$ leads to the internal boundary condition
being equivalent to
\begin{equation}
	(1-\gamma)\delta\frac{1}{2}(\ur+\ul)=\gamma\mu(\ur-\ul)\,.
	\label{Ealt}
\end{equation}
This has a pleasing symmetry as $(\ur+\ul)/2$ on the left is like the
averaging operator $\mu$ on the right, and the $(\ur-\ul)$ on the
right is like the differencing operator $\delta$ on the left.
However, at this stage I also do not know why this should be
desirable.  I leave this aspect for further research.

Approximations to fourth-order accuracy in space will be explored to
verify their accuracy.  Fourth-order approximation formulae are
obtained by truncating the analysis to errors $\Ord{\gamma^3}$.  This
enables the direct interaction between neighbouring elements through
the $\Ord{\gamma}$ terms and the indirect interaction with the next
nearest neighbours through $\Ord{\gamma^2}$ terms.  Running the
computer algebra program listed in Appendix~\ref{Sca} to higher order
in $\gamma$ gives the following for the evolution of the amplitudes
$u_j$ (using the centred difference and averaging operators $\delta$
and $\mu$ for compactness):
\begin{eqnarray}
	\dot\uj & = &
	\frac{\gamma}{h^2}\delta^2\uj
	-\frac{\gamma\alpha}{h}\uj\mu\delta\uj
	-\beta\uj^3
	+\gamma\left[ \frac{\alpha^2}{12}\uj^2\delta^2\uj
	-\frac{\beta}{24}(3\uj^2\delta^2\uj+\delta^2\uj^3) \right]
	\nonumber\\&&{}
	+\gamma^2\left[
	\frac{1}{12h^2}\delta^4\uj
	\right.\nonumber\\&&\left.\quad{}
	+\frac{\alpha}{48h}\left( 2\mu\delta(\uj\delta^2\uj)
	+24\uj\mu^3\delta\uj -10\mu\delta\uj^2 -4\uj\mu\delta\uj \right)
	\right.\nonumber\\&&\left.\quad{}
	+\frac{\alpha^2}{24}\left(\uj\delta^2\uj^2-4\uj^2\delta^2\uj\right)
	+\frac{\alpha^2}{5760}\left( 5\delta^2(\uj^2\delta^2\uj)
	-69\uj^2\delta^4\uj
	\right.	\right.\nonumber\\&&\left.\left.\qquad{}
	 -120\uj\delta^2(\uj\delta^2\uj)
	-58\uj(\delta^2\uj)^2 \right)
	\right.\nonumber\\&&\left.\quad{}
	+\frac{\beta}{8}\left( \delta^2\uj^3 -2\uj\delta^2\uj^2
	+3\uj^2\delta^2\uj \right)
	+\frac{\beta}{1440}\left( -\delta^4\uj^3
	\right.\right.\nonumber\\&&\left.\left.\qquad{}
	+87\delta^2(\uj^2\delta^2\uj) +87\uj(\delta^2\uj)^2
	+18\uj^2\delta^4\uj \right)
	\right]
	\nonumber\\&&{}
	+\Ord{\|u\|^4,\gamma^3}\,.
	\label{Eburg4}
\end{eqnarray}
The first line are terms from the model discussed in the previous
section, but here shown to lower order in $\|u\|$.  The
second line gives the appropriate correction for the linear diffusion
term $u_{xx}$ to make it fourth-order accurate in space.  The third
line gives a fourth-order correction for the advection terms $\alpha u
u_x$.  The fourth and further lines give spatially higher-order
corrections to the nonlinearly higher-order terms.

\begin{figure}[tbp]
	\centering
	\includegraphics[width=0.9\textwidth]{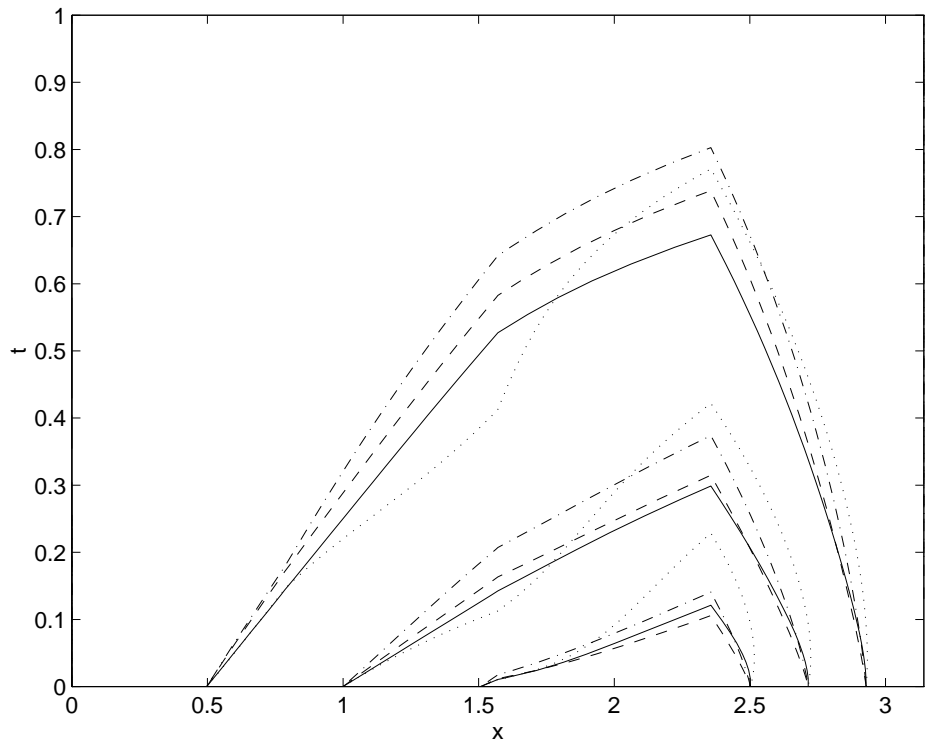}
	\caption{Contours of exact solution $u(x,t)$, -----, to compare
	with fourth-order numerical approximations~(\ref{Eburg4}):
	$\cdot\;\cdot\;\cdot\;\cdot$, the conventional approximation,
	errors $o(\alpha)$; $-\cdot-\cdot-$, the first correction,
	$o(\alpha^2)$; and $-~-~-$ includes the second correction,
	$o(\alpha^3)$.  Burger's equation~(\ref{Epde}) with parameters
	$\alpha=10$ and $\beta=0$ is discretised on $m=8$ intervals and
	drawn with contour interval $\Delta u=0.2$.}
	\label{F48a}
\end{figure}
\begin{table}[tbp]
	\centering
\caption{Comparison of the errors $\varepsilon$, defined
in~(\ref{Eerr}), of the spatially fourth order model~(\ref{Eburg4})
for the first and third truncations in the nonlinear parameter
$\alpha$ for various grid subintervals $m$.  Observe that the
higher order model is significantly less sensitive to the
nonlinearity.}
\begin{tabular}{|r|rr|rr|}
	\hline
	 & \multicolumn{2}{c|}{$m=8$}  & \multicolumn{2}{c|}{$m=16$} \\
	$\alpha$ & $o(\alpha)$ & $o(\alpha^3)$ & $o(\alpha)$ & $o(\alpha^3)$ \\
	\hline\hline
	1 & .0013 & .0016 & .0003 & .0003 \\
	\hline
	3 & .0070 & .0090 & .0009 & .0021 \\
	\hline
	6 & .0252 & .0131 & .0020 & .0052 \\
	\hline
	10 & .0464 & .0132 & .0066 & .0066 \\
	\hline
	20 & .0659 & .0171 & .0179 & .0056 \\
	\hline
	30 & .0683 & .0196 & .0250 & .0061 \\
	\hline
\end{tabular}
	\label{Ta4}
\end{table}
These fourth-order in space terms ensure a higher accuracy.  Setting
the artificial parameter $\gamma=1$ and truncating at various orders
in $\|u\|$, equivalently in $\alpha$ because I set $\beta=0$ as
before, we find that to see any appreciable difference between the
nonlinearly higher-order models we have to increase the nonlinear
parameter to $\alpha=10$ or more (from the $6$ used in the previous section).
See in Figure~\ref{F48a} that the nonlinearly low-order approximation
is a little in error, but that the higher-order models are generally
better.  They are reasonably accurate despite the large nonlinearity,
$\alpha=10$, even though there is only $m=8$ intervals in one period.
A more comprehensive summary of the numerical errors in given in
Table~\ref{Ta4} which compares the overall error $\varepsilon$ for
$m=8$ and $m=16$ elements over one period for the nonlinearity parameter
ranging over $1\leq\alpha\leq30$.  Once again observe that the
conventional $o(\alpha)$ model has errors roughly proportional to
$\alpha$, whereas the $o(\alpha^3)$ model is much less sensitive.  As
the model is fourth-order in space the errors here are an order of
magnitude smaller than those in Table~\ref{Ta2} of the previous
section.  This confirms the effectiveness of this approach to developing
finite difference models.

However, because of the subtleties in the analysis of the nonlinear
terms, the equivalent partial differential equation of a derived
finite difference model will not reduce to the original \pde{} to the
expected order in $h$.  For example, here the equivalent \pde{} of the
model~(\ref{Eburg4}), with $\gamma=1$, is
\begin{equation}
	\D tu+\alpha u\D xu=\DD xu-\beta u^3
	+\frac{h^2}{12}\left( -2\alpha u_xu_{xx}+\alpha^2uu_x^2 \right)
	+\Ord{h^4}\,,
	\label{Eequiv}
\end{equation}
which apparently differs from the generalised Burger's equation by
$\Ord{h^2}$.  But note that this equivalent differential equation is
obtained as $h\to0$ keeping $\|u\|$ fixed, whereas the centre manifold
model~(\ref{Eburg4}) is derived for fixed $h$ as $\|u\|\to 0$
but taking full account of nonlinear dynamics within the domain.  The
numerical results presented here support my contention that this
centre manifold approach better models the \pde{} for finite
$h$ and $\|u\|$.\footnote{It is conceivable that a more carefully
crafted interaction between the elements, modifying~(\ref{Esbci}), may
achieve $\Ord{h^{2p}}$ consistency between the finite difference model
and \emph{all} terms of the original \pde{} when constructing such a
centre manifold model to $\Ord{\gamma^{p+1},\|u\|^q}$.  }

\section{Conclusion}

Using centre manifold theory is a powerful new approach to deriving
finite difference models of dynamical systems.  Many details need to
researched for a general application of the theory.  However, this
particular example application to a generalised Burger's
equation~(\ref{Epde}) shows many promising features.

By introducing artificial internal boundaries we apply centre manifold
theory (\S\ref{Scm}).  The internal boundaries divide the domain of
interest into sub-domains that look very much like finite elements.
The theory guarantees essential properties required for a
low-dimensional model, though
the need to evaluate the asymptotic expressions for $\gamma=1$ weakens
this assurance.  First, a model exists parameterized in any reasonable
way we desire.  Second, the model is approached exponentially quickly
by the original dynamics, in a time of $\Ord{h^2}$ for Burger's
equation, but possibly much quicker for higher-order \pde{}s.  The
same theorem also assures us that the numerical model shares the same
stability as the original \pde{} dynamics.  Lastly, an approximation
to the model may be found to any order in the nonlinearity, using
computer algebra for example.  The resultant finite difference models
are independent of valid rearrangements of the governing \pde{}.  In
effect this technique analyses the actual dynamics of the \pde{} as a
whole---this is a holistic approach.  For example, I am not aware of
any other discretisation method that develops cross terms between the
nonlinearities, yet the presence of $\alpha\beta$ and $\alpha^2\beta$ terms
in~(\ref{Ecmg24}) suggests that they are essential for good finite
difference models.  Specific numerical simulations in \S\ref{Sbe} and
\S\ref{Shoa} show that the finite difference models derived for
Burger's equation are indeed accurate.

Further research is needed to incorporate physical boundary conditions
into the model.  The extra analysis should be straightforward, but the
level of detail would increase significantly as near the physical
boundary we would lose the translational symmetry between the elements.
Then the intriguing issue of initial conditions needs further work.
Following \cite{Roberts89b,Cox93b,Roberts97b} we note that to make
accurate forecasts with the numerical models we need to provide
initial values $\uj^0$ which are not the initial field $u_0(x)$
sampled at the collocation points $x_j$.  Instead preliminary work
shows that $\uj^0$ should be approximately the average of $u_0(x)$
over each element.  Non-autonomous forcing of the \pde{} will need
projecting onto the finite difference grid in a similar manner.
The
analysis of other \pde{}s in possibly higher-spatial dimensions would
appear to hinge upon being able to solve the linear part of the \pde{}
over the elements subject to forcing from the coupling across the
artificially introduced internal boundaries.  This is potentially
quite complicated, but the results of the simple problems here suggest
that the analysis could be very worthwhile.

\appendix
\section{Computer algebra develops the approximations}
\label{Sca}

Just one of the virtues of this centre manifold approach to modelling
is that it is systematic.  This enables relatively straightforward
computer programs to be written to find the centre manifold and the
evolution thereon \cite[e.g.]{Roberts96a}.  I believe computer
algebra will be increasingly used to support research by performing
extensive routine algebraic manipulations.  To ensure correctness and
to provide a basis for further work it seems reasonable to include
the computer algebra code.

For this problem the iterative algorithm is implemented by a
computer algebra program written in \textsc{reduce} \footnote{At
the time of writing, information about \texttt{reduce} was
available from Anthony C.~Hearn, RAND, Santa Monica,
CA~90407-2138, USA. E-mail: \tt reduce@rand.org} Although there
are many details in the program, the correctness of the results
are \emph{only determined} by driving to zero (line~59) the
residual of the governing differential equation, evaluated on
line~48, to the error specified on line~46 and with the residual
of the internal boundary condition computed on lines~50--53.
The other details only affect the rate of convergence to the
ultimate answer.

{\footnotesize \verbatimlisting{holistic.red} }

\addcontentsline{toc}{section}{\refname}
\markright{\textnormal{\sf\refname}}
\bibliographystyle{plain}\bibliography{ajr,bib,new}

\end{document}